\documentclass{IEEEtran4PSCC}

\ifCLASSINFOpdf
   \usepackage[pdftex]{graphicx}
\else
   \usepackage[dvips]{graphicx}
\fi

\usepackage[cmex10]{amsmath}

\makeatletter
\let\old@ps@headings\ps@headings
\let\old@ps@IEEEtitlepagestyle\ps@IEEEtitlepagestyle
\def\psccfooter#1{%
    \def\ps@headings{%
        \old@ps@headings%
        \def\@oddfoot{\strut\hfill#1\hfill\strut}%
        \def\@evenfoot{\strut\hfill#1\hfill\strut}%
    }%
    \def\ps@IEEEtitlepagestyle{%
        \old@ps@IEEEtitlepagestyle%
        \def\@oddfoot{\strut\hfill#1\hfill\strut}%
        \def\@evenfoot{\strut\hfill#1\hfill\strut}%
    }%

    \ps@headings%
}
\makeatother

\usepackage{mathrsfs}%
\usepackage{amsmath,amsfonts}
\usepackage{algorithm}
 \usepackage{algpseudocode}
\usepackage{array}
\usepackage{textcomp}
\usepackage{stfloats}
\usepackage{subfig}
\usepackage{float}
\usepackage{url}
\usepackage{verbatim}
\usepackage{graphicx}
\usepackage{cite}
\usepackage{eurosym}

\usepackage{multirow}
\usepackage{hyperref}
\usepackage{enumitem}
\usepackage{graphicx}
\usepackage{nomencl}
\usepackage[compact]{titlesec}         
\titlespacing{\section}{0pt}{0pt}{0pt} 
\AtBeginDocument{
  \setlength\abovedisplayskip{0pt}
  \setlength\belowdisplayskip{0pt}}
  
\usepackage[table]{xcolor} 
\makenomenclature
\usepackage{tikz}
\usepackage{adjustbox}
\usepackage{mathtools}
\usepackage{nomencl}
\makenomenclature

\begin{document}

\setlength{\abovedisplayskip}{5pt}
\setlength{\belowdisplayskip}{5pt}

\title{Feature-Driven Strategies for Trading \\  Wind Power and Hydrogen}

\author{Emil Helgren, Jalal Kazempour, and Lesia Mitridati \\
    Department of Wind and Energy Systems, Technical University of Denmark, Kgs. Lyngby, Denmark\\
    helgrenemil@gmail.com, \{jalal, lemitri\}@dtu.dk \\ }




\maketitle


\begin{abstract}
This paper develops a feature-driven model for hybrid power plants, enabling them to exploit available contextual information such as historical forecasts of wind power, and make optimal wind power and hydrogen trading decisions in the day-ahead stage. For that, we develop different variations of feature-driven linear policies, including a variation where policies depend on price domains, resulting in a price-quantity bidding curve. In addition, we propose a real-time adjustment strategy for hydrogen production. Our numerical results show that the final profit obtained from our proposed feature-driven trading mechanism in the day-ahead stage together with the real-time adjustment strategy is very close to that in an ideal benchmark with perfect information. 
\end{abstract}


\vspace{5mm}
\section{Introduction}
Hybrid power plants, comprising an electrolyzer and wind turbines, are expected to be largely installed in many countries, such as Denmark \cite{PtX_strategy_DK}. The plant operator, aiming to maximize their profit, can either allocate the entire generated wind power for hydrogen production, sell the whole or part of generated wind power to the grid, or buy power from the grid to produce more hydrogen. The plant operator should make informed decisions for trading power and hydrogen in a forward stage, when neither day-ahead and balancing prices nor the true wind power generation are realized. This requires developing a trading strategy, which is the focus of this paper.

There are many papers in the literature that develop trading strategies for wind power producers (wind power as the sole product) under price and wind uncertainty using stochastic optimization methods. Among others, \cite{pinson2007trading} solves a newsvendor problem, \cite{Morales2010} develops a two-stage stochastic programming, \cite{xu} devises a range of robust optimization models, and finally \cite{pinsondro} develops a distributionally robust model. In addition, various data-driven approaches have been proposed that take advantage of contextual information to train a probabilistic forecast model and use its outcomes to solve a stochastic optimization problem, as summarized in \cite{sadana2023survey}. These approaches include sequential  training and optimization methods, in which the probabilistic forecast model is trained independently of the optimization task \cite{bertsimas2019}, and integrated training and optimization methods, in which the forecast model is trained in order to minimize the losses in a specific optimization task \cite{elmachtoub2022smart}.
While all these methods outperform the deterministic counterpart, they require (\textit{i}) knowledge of probabilistic forecasting, (\textit{ii}) generating a set of scenarios, an uncertainty set, or a family of probability distributions, and then (\textit{iii}) developing a stochastic optimization model to be solved in the decision-making stage. The trading problem is even more challenging for hybrid power plants, where hydrogen is also a product in addition to wind power. Although one may expect large companies to have great expertise in making trading decisions under uncertainty, it might be more challenging for a small-scale hybrid power plant operator. Therefore, the aim of this paper is to develop a \textit{pragmatic} trading approach, which outperforms the deterministic model by enabling the hybrid power plant to learn from historical data, without the need for the plant to use probabilistic forecasting and complex stochastic solutions.

The first contribution of this paper is to propose a novel application of a prescriptive analytics framework based on decision rules, inspired by \cite{liyanage2005practical}, to the multi-market bidding problem of hybrid power plants. In this data-driven approach, we exploit contextual information, so-called \textit{features}, such as historical (deterministic) forecasts of wind power and directly map them to an optimal action. In the \textit{training stage}, the hybrid power plant operator learns feature-driven \textit{policies} based on historical features and uncertainty realizations. In the \textit{decision-making stage}, the learned policies are applied to new available features, leading to trading decisions without the need to solve a complex optimization problem. A few papers in the literature, e.g., \cite{munoz2020feature}, \cite{stratigakos2021prescriptive} and \cite{Parginos}, develop feature-driven trading models for renewable power producers. To the best of our knowledge, this is the first paper that develops such a model for hybrid power plants trading both wind power and hydrogen, which is a more complicated problem. The second contribution of this paper is to extend these previous works by investigating various model architectures, in particular price- and time-dependent policies, and feature vectors, including an improved forecast feature vector. Finally, the third contribution of this paper is to develop a pragmatic rule-based adjustment strategy for real-time hydrogen production. Using an out-of-sample simulation, we show how the resulting profit from feature-driven trading in the day-ahead stage and the adjustment strategy in real time is very close to that in an ideal benchmark (oracle).


The rest of the paper is organized as follows. Section \ref{sec:decision_making_framework} explains the decision-making framework. Section \ref{sec:decision_making_framework} details the feature-driven trading strategy model in the day-ahead stage. Section \ref{sec:adjustment_task_in_rt} presents the proposed real-time adjustment strategy. Section \ref{sec:case_study} provides numerical results. Finally, Section \ref{sec:conclusion} concludes the paper.  

\vspace{5mm}
\section{Decision-Making Framework}\label{sec:decision_making_framework}




We consider a hybrid power plant comprising an electrolyzer and a wind turbine, behind the meter. The wind power generation can be exported to the grid or directed to the electrolyzer to produce hydrogen with a constant efficiency coefficient\footnote{The efficiency of electrolyzers in practice is variant, depending on their power consumption level following a non-linear curve.  We refer the interested reader to \cite{conic} for alternative convexified solutions.}. The electrolyzer has also the possibility to buy power directly from the grid\footnote{According to the ‘‘Renewable Fuels of Non-Biological Origin (RFNBO)’’
Delegated Act of the European Commission published in 2023 \cite{RFNBO}, the produced hydrogen in this case will not be seen green except for certain operational conditions. We do not consider this legislation in our paper.}. The produced hydrogen is sold through a bilateral contract at a fixed hydrogen price. As part of the contract, there is a minimum amount of hydrogen that should be produced in a daily basis. In practice, this minimum daily requirement is fulfilled by filling tube trailers that are collected daily by the hydrogen off-taker. However, the proposed models could be generalized straightforwardly to other off-take structures, such as injection of hydrogen in gas pipelines. The hydrogen production thus needs to be scheduled for the whole day in order to ensure fulfilling this minimum production requirement. Any real-time adjustment should likewise account for the imposed hydrogen production quota. The power production capacity of the wind farm is assumed to be identical to the power consumption capacity of the electrolyzer.

The hybrid power plant aims to maximize the total profit from electricity trading and hydrogen sales. This leads to two stages for decision making:

\textit{Day-ahead decision making}: Given the available (deterministic) wind power forecast, the hybrid power plant must decide how much electricity to trade (buy or sell) in the day-ahead market and how to schedule the operation of the electrolyzer for each hour of the following day, depending on hourly day-ahead market prices and the fixed hydrogen price. Recall the minimum daily hydrogen production requirement needs to be accounted for. In Nord Pool, all these decisions for every hour $t \in \mathcal{T}$ of day $\rm{D}$ should be made before noon of day $\rm{D}\!-\!1$. 

\textit{Real-time decision making}: The hybrid power plant must settle imbalances arising from deviations in wind power production (compared to the day-ahead schedule) in the balancing market. Therefore, at the time of delivery, i.e., in every hour $t$ of day $\rm{D}$, the plant must decide how to adjust the schedule of the electrolyzer based on the realized wind power production and electricity balancing prices, while ensuring the  minimum daily hydrogen production requirement is fulfilled.



\vspace{5mm}
\section{Day-Ahead Bidding}\label{sec:learning_task_for_da_bidding
}
We first define a feature-driven model for making power and hydrogen trading decisions in the day-ahead stage. Next, we explain  feature vectors, and then elaborate on architecture of linear policies. We eventually provide an optimization model for training the proposed feature-driven model.

\vspace{2mm}
\subsection{Model Definition} \label{Architecture}
\vspace{-1mm}
The main challenge for the hybrid power plant in the day-ahead stage is to account for uncertainty on day-ahead and balancing electricity prices and wind power production. 

Aiming to develop a feature-driven trading strategy in the day-ahead stage, we first define the vector of available features with $N$ elements corresponding to hour $t$ as 
%
\begin{align}
&\textbf{X}_t = \Big[ X^1_t, ..., X^{N}_t\Big] \ \ \forall t \in \mathcal{T}.
\end{align}

The features should be those contextual information for hour $t$ of the following day that the plant has access to when making day-ahead decisions. Examples of available features for the hybrid power plant at the day-stage stage are (\textit{i}) deterministic forecasts of wind power generation of the underlying farm at different points before the day-ahead market gate closure at the noon of day $\rm{D}\!-\!1$, (\textit{ii}) aggregate wind power forecast in the area (or bidding zone) where the farm is located, (\textit{iii}) aggregate wind power forecast in the neighboring areas, (\textit{iv}) realized wind power production in day $\rm{D}\!-\!1$.

Let variable $p^{\rm DA}_t$ denote the electricity traded (bought or sold) in the day-ahead market for hour $t$ of the next day. In addition, variable $p^{\rm H}_t$ is the electricity consumption schedule of the electrolyzer for the same hour. We exploit linear decision rules \cite{kuhn}, so called \textit{linear policies}, to define $p^{\rm DA}_t$ and $p^{\rm H}_t$ as a function of features $\textbf{X}_t$, such that
%
%
\begin{subequations} \label{eq:def}
\begin{align}
    & p^{\rm DA}_t = \textbf{q}^{\rm{DA}} \ \textbf{X}^\top_t &  \forall t \in \mathcal{T} \label{eq:def_q_DA} \\
    & p^{\rm H}_t = \textbf{q}^{\rm{H}} \ \textbf{X}^\top_t &  \forall t \in \mathcal{T}, \label{eq:def_q_H}  
\end{align}
\end{subequations}
where $(.)^{\top}$ is the transpose operator. Vectors $\textbf{q}^{\rm{DA}} \in \mathbb{R}^{N}$ and $\textbf{q}^{\rm{H}} \in \mathbb{R}^{N}$ are the policies, so called $\textbf{q}$-policies, to be learned from historical data in the training stage. 

If the model is designed and trained appropriately, $\textbf{q}^{\rm{DA}}$ and $\textbf{q}^{\rm{H}}$ will reflect relations between the input features and the decision variables $p^{\rm DA}_t$ and $p^{\rm H}_t$ that persist into the future, such that the model will produce well-performing trading decisions for new feature vectors. 

\begin{figure}[t]
    \centering
    \includegraphics[width=0.75\linewidth]{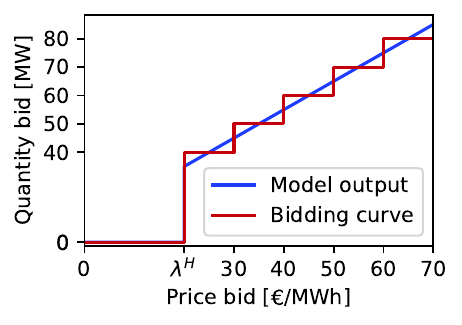}
    \caption{The discretization of $p^{\rm DA}_t=$ $ \textbf{q}^{{\rm{DA}}} \textbf{X}^\top_t$ to create piece-wise price-quantity bids $\{\lambda_b,p_b\}_{b=1, ..., B}$.} 
    \label{fig:03_discretizing}
\end{figure}

\textit{Feature Vectors}:
The day-ahead market allows for simultaneously placing multiple independent price-quantity bids $\{\lambda_b,p_b\}_{b=1, ..., B}$ for each hour of the following day. Therefore, we can model the electricity traded in the day-ahead market $p^{\rm DA}_t$ as a function of the day-ahead price at hour $t$.
This is achieved by introducing an unknown price variable $\lambda\!\in\!\mathbb{R}$ in the feature vector $\textbf{X}_t$. Since all other values in the feature vector are known at the time of decisions, this results in $p^{\rm DA}_t$ being a linear function of this unknown variable $\lambda$. In practice, the day-ahead price-quantity bids are then derived by discretizing this linear function and approximating it as a stepwise function with arbitrary step size, as illustrated in Fig. \ref{fig:03_discretizing}.
Each step $b$ of the resulting stepwise function will represent one pair of price-quantity bids $\{\lambda_b,p_b\}_{b=1, ..., B}$. Since the electricity traded in the day-ahead market and the power consumption $p^{\rm H}_t$ of the electrolyzer are interdependent, $p^{\rm H}_t$ is also expressed as a linear function of the day-ahead electricity price at hour $t$. 

All feature vectors investigated in this paper will have the form 
$\textbf{X}_t = \Big[ \tilde{\textbf{X}}_t, \lambda, 1 \Big]$, where the constant feature $1$ provides an intercept and $\tilde{\textbf{X}}_t \in \mathbb{R}^{N-2}$ is the vector of remaining features. The remaining features can include any relevant data that are expected to be related with the uncertain parameters including the wind power production (e.g., wind power forecasts) and the day-ahead and balancing prices (e.g., electricity price and demand forecasts and imbalance forecasts). Note that the plant has usually access to an extensive database, including \textit{historical} values of these features and corresponding \textit{realized} values of the uncertain parameters.


\textit{Architecture of Linear Policies}: 
Several approaches can be used regarding the architecture of the $\textbf{q}$-policies. The simplest but restrictive architecture, so-called \textit{general architecture}, considers \textit{constant} policies $\textbf{q}^{\rm DA} \in \mathbb{R}^{N}$ and $\textbf{q}^{\rm H} \in \mathbb{R}^{N}$ for every hour $t$.

It is widely acknowledged that daily patterns exist in the day-ahead prices. Accounting for these patterns could remarkably increase the performance of the model. This can be achieved by introducing an \textit{hourly architecture}, where the $\textbf{q}$-policies in \eqref{eq:def} are specific to the hour of the day and denoted as $\textbf{q}^{\rm DA}_{h_t} \in \mathbb{R}^{N}$ and $\textbf{q}^{\rm H}_{h_t} \in \mathbb{R}^{N}$ for $h_t = t \mod 24$, with $\mod$ representing the modulo operator. This hourly architecture allows the model to learn daily patterns. However,  it results in $24$ times as many $\textbf{q}$-policies to train as in the general architecture, and thus requires a larger training dataset. The trade-off between these factors will be empirically studied later in our case study.

Besides, in order to capture more complex and non-linear dependencies between the features $\textbf{X}_t$ and decision variables $p^{\rm DA}_t$ and $p^{\rm H}_t$, multiple sets of model parameters can be introduced, each covering one domain of the features. In particular, we introduce $M$ price domains $\mathcal{D}\!=\!\{\mathcal{D}_1,...,\mathcal{D}_M\}$, each defined as a range of prices $\mathcal{D}_i=\big[\lambda_{i-1},\lambda_{i}\big]$ with $\lambda_{i-1}<\lambda_{i}$ for all $i \in \{1, ..., M\}$.
Over each price domain, we define $\textbf{q}$-policies as $\textbf{q}^{\rm DA}_{i}$ and $\textbf{q}^{\rm H}_{i}$.
As a result, the decision variables $p^{\rm DA}_t$ and $p^{\rm H}_t$ are expressed as piece-wise linear functions of the day-ahead prices, such that
\begin{subequations}
    \begin{align}
        & p^{\rm DA}_t = \textbf{q}^{\rm DA}_{i} \ \textbf{X}^\top_t  \hspace{-5mm}  & \text{ if } \lambda \in \mathcal{D}_i, \hspace{1cm} & \forall t \in \mathcal{T} \\
        & p^{\rm H}_t = \textbf{q}^{\rm H}_{i} \ \textbf{X}^\top_t  \hspace{-5mm} & \text{ if } \lambda \in \mathcal{D}_i, \hspace{1cm} & \forall t \in \mathcal{T}.
    \end{align}
\end{subequations}

Note that a trade-off between complexity and flexibility can be achieved, if optimal domains can be chosen based on statistical characteristics of the data. For example, a natural threshold where it might be beneficial to divide the model into different trading strategies is when the day-ahead electricity price is equal to the hydrogen price $\lambda^{\rm H}$.

\vspace{2mm}
\subsection{Optimization Model for Training Stage}\label{sec:DA_training}
\vspace{-1mm}
In the training stage we determine the adequate values of the $\textbf{q}$-policies using a batch learning mechanism. The training of the $\textbf{q}$-policies can be represented as an optimization problem, in which the profit of the plant is maximized over a \textit{historical} dataset, containing values of the feature vectors along with corresponding realizations of the uncertain parameters, e.g., market prices, for all $t \in \mathcal{T}^{\rm hist}$. This optimization problem, formulated as a mixed-integer linear program (MILP), reads as
\begingroup
\allowdisplaybreaks
\begin{subequations} \label{opt}
 \begin{alignat}{3}
    & \max_{\Omega} &&  \sum_{t \in \mathcal{T}^{\rm hist}} \Big[\lambda^{\rm DA}_t \textbf{q}^{\rm{DA}}  \textbf{X}^\top_t + \lambda^{\rm H} \rho^{\rm H}\textbf{q}^{\rm{H}}  \textbf{X}^\top_t && +  \lambda^{\rm UP}_t o_t - \lambda^{\rm DW}_t u_t  \Big]  \label{eq:training_1} \\
    & \text{s.t.} && o_t  - u_t = P^{\rm W}_t - \textbf{q}^{\rm{DA}}  \textbf{X}^\top_t - \textbf{q}^{\rm{H}}  \textbf{X}^\top_t   && \hspace{2.5mm} \forall t \in \mathcal{T}^{\rm hist} \label{eq:training_2} \\ 
    &   \quad &&  0 \leq o_t \leq M  (1 - b_t)   && \hspace{2.5mm} \forall t \in \mathcal{T}^{\rm hist} \label{eq:training_7} \\
    &  \quad &&  0 \leq u_t \leq M  b_t   && \hspace{2.5mm} \forall t \in \mathcal{T}^{\rm hist} \label{eq:training_8} \\
    &  \quad && 0 \leq \textbf{q}^{\rm{H}}  \textbf{X}^\top_t \leq \overline{P}^{\rm H}   && \hspace{2.5mm} \forall t \in \mathcal{T}^{\rm hist} \label{eq:training_9} \\
    &  \quad &&  -\overline{P}^{\rm H} \leq \textbf{q}^{\rm{DA}}  \textbf{X}^\top_t \leq \overline{P}^{\rm W}   && \hspace{2.5mm} \forall t \in \mathcal{T}^{\rm hist} \label{eq:training_10} \\
    &  \quad && \sum_{t=24(d-1)+1}^{24d} \rho^{\rm H}\textbf{q}^{\rm{H}}  \textbf{X}^\top_t \geq \underline{H}   && \hspace{2.5mm} \forall d \in D^{\rm hist}\label{eq:training_11},
\end{alignat}   
\end{subequations}
\endgroup
 where the set of variables $\Omega\!=\!\{\textbf{q}^{\rm DA}$, $\textbf{q}^{\rm H}$, $o_t$, $u_t$, $b_t\} \ \forall{t} \in \mathcal{T}^{\rm hist}$ includes the $\textbf{q}$-policies $\textbf{q}^{\rm DA}$ and $\textbf{q}^{\rm H}$, as well as the real-time decisions, including the over-production $o_t$ and under-production $u_t$ settled in real time, and the auxiliary binary variable $b_t\!\in\!\{0,1\}$ indicating the state of over- or under-production. Note that the $\textbf{q}$-policies in \eqref{opt} can take index $h_t$  and/or index $i$ as discussed in Section \ref{Architecture} 

The objective function \eqref{eq:training_1}  maximizes the total profit of the hybrid power plant over the historical dataset, including four terms. The first term $\lambda^{\rm DA}_t \textbf{q}^{\rm{DA}}  \textbf{X}^\top_t$ corresponds to the revenue/cost from electricity trading in the day-ahead market, where $\lambda_t^{\rm DA}\!\in\!\mathbb{R}$ is the historical day-ahead price reported by the market operator. Note that $\textbf{q}^{\rm{DA}}  \textbf{X}^\top_t$, reflecting the power quantity, could be positive or negative, indicating whether the plant sells or buys power. The second term  $\lambda^{\rm H} \rho^{\rm H}\textbf{q}^{\rm{H}} \textbf{X}^\top_t$ is the revenue from hydrogen sales, where $\lambda^{\rm H}\!\in\!\mathbb{R}_+$, in €/Kg, is the fixed hydrogen price. In addition, $\rho^{\rm H}\!\in\!\mathbb{R}_+$, in Kg/MWh, is the constant power-to-hydrogen efficiency of the electrolyzer, whereas $\textbf{q}^{\rm{H}}  \textbf{X}^\top_t$ reflects the power consumed by the electrolyzer. Finally, the third and fourth terms in \eqref{eq:training_1} refer to the settlement of over- and under-production, respectively, in the balancing market with a dual-price imbalance settlement scheme. Note that $\lambda_t^{\rm UP}\!\in\!\mathbb{R}$ is the upward regulation price, which is lower than or equal to the day-ahead price $\lambda_t^{\rm DA}$, reflecting a lost opportunity cost for over-production. In addition, $\lambda_t^{\rm DW}\!\in\!\mathbb{R}$ is the downward regulation price, which is higher than or equal to the day-ahead price $\lambda_t^{\rm DA}$, reflecting a penalty cost for under-production. By this, it is always costly for the hybrid power plant to create any imbalance either  over- or under-production.


Constraint \eqref{eq:training_2} defines the power imbalance $o_t\!-\!u_t$ to be settled in real time as the difference between the realized wind power generation $P^{\rm W}_t$ and the power scheduled in the day-ahead stage. The set of disjunctive constraints \eqref{eq:training_7}-\eqref{eq:training_8}, with $M$ a large enough positive constant, enforces over- and under-production to not happen simultaneously at each hour $t$. If $b_t = 1$, then \eqref{eq:training_7} enforces the over-production $o_t$ to be zero. On the other hand, if $b_t = 0$, then \eqref{eq:training_8} sets the under-production $u_t$ to zero. Constraint \eqref{eq:training_9} enforces the power consumption of the electrolyzer to lie within zero and its capacity $\overline{P}^{\rm H}$. Note that the lower bound being set to zero means that the electrolyzer cannot operate as a fuel cell, converting the hydrogen back to electricity. In order to avoid imbalance settlement costs, the power traded in the day-ahead market is restricted by \eqref{eq:training_10} to lie within the minus consumption capacity of the electrolyzer  and the nominal capacity of the wind turbine $\overline{P}^{\rm W}$.
Finally, the minimum daily hydrogen production requirement $\underline{H}$ is enforced by \eqref{eq:training_11} for each day $d \in D^{\rm hist}$ in the historical dataset. This requirement could be alternatively implemented using a  slack variable penalized in the objective function. However, a value for such a penalty is unknown. One may assume the penalty would always be higher than the revenue that could be generated by not meeting the requirement.

\vspace{2mm}
\subsection{Decision-Making Stage and Retraining}
\vspace{-1mm}
For each day of the testing period, the trained $\textbf{q}$-policies are applied to a new feature vector $\tilde{\textbf{X}}_t$ for each hour $t$ of the following day, to compute the linear functions representing the electricity traded in the day-ahead market and consumed by the electrolyzer as functions of the future day-ahead prices. These functions are then discretized, and stepwise price-quantity bids are placed in the day-ahead market. The realized day-ahead price will then determine the electricity traded in the day-ahead market and consumed by the electrolyzer. For the sake of simplicity, we assume none of the technical constraints of the hybrid power plant is violated in the testing period. This is in practice likely as long as the testing data does not differ significantly from the training one. If it is the case, one should restore feasibility of those constraints by adjusting trading decisions at the minimum cost, or projecting the decisions onto the feasible space.

\begin{figure}[t]
    \centering
    \includegraphics[width=0.7\linewidth]{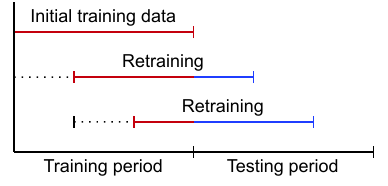}
    \caption{Illustration of sliding window approach to determine training data in each retraining interval. 
    }
    \label{fig:03_sliding_window}
\end{figure}

Once the uncertain parameters, i.e., day-ahead prices, balancing prices, and wind power production, are realized, the new feature vector and the associated realizations are added to the historical dataset. Retraining the model considering more recent historical data points is thus performed at regular intervals. One can use a sliding window approach in order to keep the length of the training dataset constant, as illustrated in Fig. \ref{fig:03_sliding_window}. Note that, at each retraining interval, the blue part of the window represents the previous testing data points recently added to the training dataset. The length of this window is a hyperparameter of the model that should be selected appropriately, as further discussed in our numerical analysis.

\vspace{6mm}
\section{Real-Time Adjustments}\label{sec:adjustment_task_in_rt}
We assume the electrolyzer is subject to negligible ramping constraints. Therefore, the hybrid power plant is able to adjust the hydrogen production schedule in real time based on the realized wind power production and balancing prices. This section introduces a rule-based adjustment algorithm for the real-time operation of the electrolyzer that accounts for the minimum daily hydrogen production requirement.

\vspace{2mm}
\subsection{Decision Rule for a Single Hour}
\vspace{-1mm}
In each hour, the decision rule is dependent on whether  balancing prices are higher or lower than the hydrogen price. Hereafter, to make electricity and hydrogen prices comparable in terms of €/MWh, we refer to $\rho^{\rm H}\lambda^{\rm H}$ as the hydrogen price. 
Balancing prices are unknown until after the balancing settlements, however it is assumed that they can be approximated to a sufficient degree for the current hour by predicting the system status \cite{wind_trading_2} and observing the intraday market. This  will be explained in further detail after the introduction of the algorithm and required parameters. 

As mentioned earlier, we assume that the balancing market follows a dual-pricing mechanism, implying that $\lambda^{\rm UP}_t \leq \lambda^{\rm DA}_t \leq \lambda^{\rm DW}_t$ for each hour $t \in \mathcal{T}$. The following three situations can thus occur regarding the relation between the hydrogen price and balancing prices: (\textit{i})  $\rho^{\rm H}\lambda^{\rm H} \leq \lambda^{\rm UP}_t \leq \lambda^{\rm DW}_t$; (\textit{ii}) $\lambda^{\rm UP}_t \leq \lambda^{\rm DW}_t \leq \rho^{\rm H}\lambda^{\rm H}$; and (\textit{iii}) $\lambda^{\rm UP}_t \leq \rho^{\rm H}\lambda^{\rm H} \leq \lambda^{\rm DW}_t$.
In the first situation where both up and down balancing prices are higher than the hydrogen price, producing hydrogen would result in less revenue than what is compensated for over-production in the balancing market. This means that the producer would benefit from adjusting the electrolyzer's consumption down and exporting as much electricity to the grid as possible. In the second situation where both balancing prices are lower than the hydrogen price, producing hydrogen results in higher revenue than the cost incurred from under-production in the balancing market. This means that the hybrid power plant would benefit from adjusting the electrolyzer's consumption up and importing as much electricity from the grid as possible. In the final situation, over-production results in lower revenues than hydrogen production, and under-production results in higher costs than revenues from hydrogen production. In this situation, the electrolyzer should be adjusted in order to minimize any deviation from the day-ahead market bid. This decision rule that maximizes the profit for a single hour can be formulated as a piece-wise linear function $\Pi(.)$ that returns the adjusted electricity consumption level of the electrolyzer, such that
\begin{align}
    \Pi \big(\delta_t, \lambda^{\rm H}, \lambda^{\rm UP}_t, \lambda^{\rm DW}_t\big) = \begin{cases}
        0 & \text { if } \lambda^{\rm UP}_t > \rho^{\rm H} \lambda^{\rm H} \\
        \overline{P}^{\rm H} & \text { if } \lambda^{\rm DW}_t < \rho^{\rm H}\lambda^{\rm H} \\
        \alpha^{\rm H}[\delta_t] & \text{ otherwise,}
    \end{cases} \label{eq:adj_rule}
\end{align}
where $\delta_t\!=\!P^{\rm W}_t\!-\!p^{\rm DA}_t$ is the surplus/deficit of wind power production compared to the day-ahead market bid, and $\alpha^{\rm H}[\cdot]$ ensures that the electrolyzer production is feasible:
\begin{equation}
    \alpha^{\rm H}\big[\delta_t\big] = 
    \begin{cases}
        0 &  \text { if } \delta_t < 0 \\
        \overline{P}^{\rm H} &  \text { if } \delta_t > \overline{P}^{\rm H} \\
        \delta_t & \text{ otherwise.}
    \end{cases} 
\end{equation}

Adjusting the electricity consumption of the electrolyzer according to the decision rule \eqref{eq:adj_rule} will maximize the generated revenues for a single hour, when the estimated balancing price is the only source of uncertainty. 
Note that \eqref{eq:adj_rule} does not require an accurate estimation of balancing prices, but only whether  balancing prices will be higher or lower than the hydrogen price, which is a more straightforward task. Whether the balancing price will be higher or lower than the known day-ahead price will depend on the system status (surplus or deficit energy in the system). 
    %
        %


\vspace{2mm}
\subsection{Real-time Adjustment Algorithm} \label{adjus}
\vspace{-1mm}
Implementing the hourly decision rule \eqref{eq:adj_rule} independently for each single hour of the optimization period would not guarantee that the minimum daily hydrogen production requirement $\underline{H}$ is satisfied. Instead, we introduce an algorithm which only adjusts the electrolyzer's schedule according to the hourly decision rule in a given hour $t$ if this requirement can still be met during the day. Therefore, the electrolyzer's consumption can always be adjusted upward from the day-ahead schedule at any given hour. However, in order to perform a downward adjustment in a given hour $t$ of the day $d$, the adjusted hydrogen production in this hour $\rho^{\rm H} p^{\rm adj}_t$, plus the cumulative realized production in previous hours $\rho^{\rm H} p_t^{\rm real} = \sum_{i=24(d-1)+1}^{t-1} \rho^{\rm H} p^{\rm adj}_i$, plus the cumulative production schedule for the remaining hours of the day $\sum_{i=t+1}^{24d} \rho^{\rm H} p^{\rm H}_i$ must be higher than the minimum daily hydrogen production requirement $\underline{H}$. For example, let us consider a hybrid power plant with a certain minimum daily hydrogen production requirement, equivalent to the consumption of $15$ MWh. If the realized cumulative hydrogen production at hour $15$ is equal to $10$ MWh and the cumulative planned hydrogen production from hour $15$ to $23$ is equal to $7$ MWh, then the electrolyzer has a scheduled hydrogen surplus of $2$ MWh, which can be adjusted down at hour $15$. If the hourly decision rule \eqref{eq:adj_rule} outputs a lower set-point for the electrolyzer than its schedule for this hour, the electrolyzer can then be turned down as much as the surplus allows. On the contrary, if the realized cumulative hydrogen production is only equal to $8$ MWh, then the electrolyzer has no scheduled hydrogen surplus and cannot be adjusted down in this hour, regardless of the outputs of the hourly decision rule \eqref{eq:adj_rule}. The size of the production requirement compared to the production capacity of both the wind farm and the electrolyzer thus directly affects the possible space of adjustments, with a higher requirement reducing the amount of adjustment possible. The complete adjustment algorithm is provided in Algorithm \ref{alg:adj_up_and_dw}.

\begin{algorithm}[t]
\caption{Real-time upward and downward adjustment}\label{alg:adj_up_and_dw}
\begin{algorithmic}[1]
\For{each day $d$}
    \State Initialize $p^{\rm real}_{24(d-1)+1} = 0$
    \For{each hour $t \in [24(d-1)
    +1, 24d]$}
        \State Receive model output: $p^{\rm DA}_t$, $\textbf{p}^{\rm H}$
        \State Receive balancing price forecasts: $\lambda^{\rm UP}_t$, $\lambda^{\rm DW}_t$
        \State Realize production: $P^{\rm W}_t$
        \State Compute real-time deviation $\delta_t = P^{\rm W}_t - p^{\rm DA}_t$
        \State Compute  $p^{\rm H*}_t = \Pi\big(\delta_t, \lambda^{\rm UP}_t, \lambda^{\rm DW}_t\big)$
        \If{$p^{\rm H*}_t > p^{\rm H}_t$}
            \State $p^{\rm adj}_t = p^{\rm H*}_t$
        \Else
            \State $p^{\rm adj}_t = \text{max}\left(
            p^{\rm H*}_t,
            \dfrac{\underline{H}}{\rho^{\rm H}} -  \left( p^{\rm real}_t + \sum\limits_{i=t+1}^{24d} p^{\rm H}_i \right)
            \right)$
        \EndIf
        \State \text{Receive revenue from allocations $\big(p^{\rm DA}_t, p^{\rm adj}_t\big)$} 
        \State $p^{\rm real}_{t+1} = p^{\rm real}_{t} + p^{\rm adj}_t$
    \EndFor
\EndFor
\end{algorithmic}
\end{algorithm}
\vspace{-2mm}
\vspace{6mm}
\section{Numerical Analysis}\label{sec:case_study}

This section provides a numerical analysis of the proposed feature-driven day-ahead trading strategy and rule-based real-time adjustment. All source codes are  available in \cite{Github}.

\vspace{2mm}
\subsection{Case Study Data}
\vspace{-1mm}
We use historical data for the year $2019$ for training, and the year $2020$ for testing, ensuring that the models are evaluated through all seasonal variations. It is essential to evaluate the models with out-of-sample testing in this way, using more recent data for testing than was used for training, to ensure the performance of the models is due to them generalizing observed patterns in the features and not overfitting the training data. All price data is publicly available on ENTSO-e \cite{entso}, along with production data for specific wind farms, and aggregated forecasts of offshore/onshore wind production in the Danish bidding zones DK1 and DK2. The production from the wind farm in Roedsand, Denmark is used to model the realized wind production of the hybrid power plant.

\begin{figure}[t]
    \centering
\includegraphics[width=\linewidth]{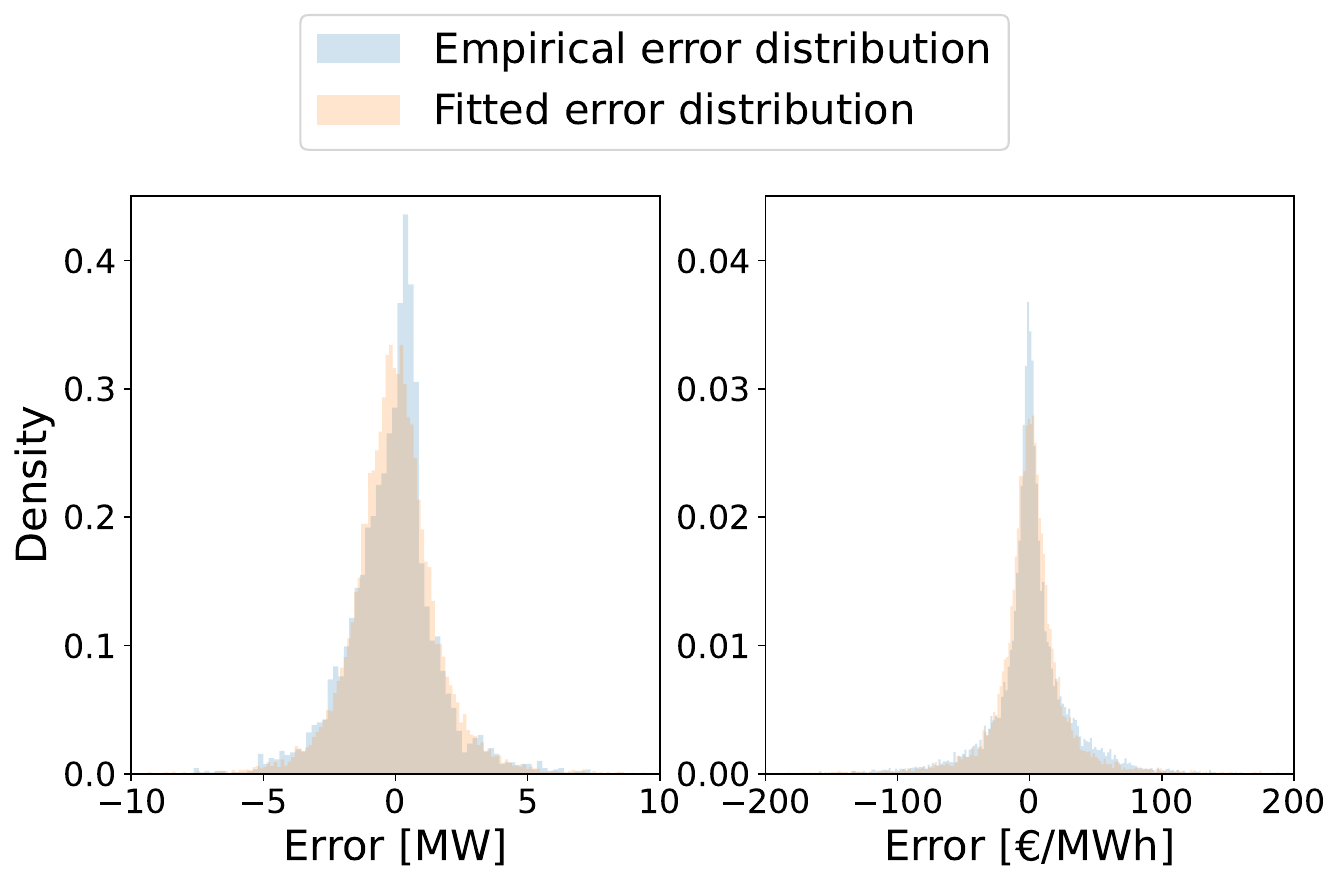}
\caption{Distribution of day-ahead market price \textbf{(right)} and wind power generation \textbf{(left)} forecast errors.}
\label{fig:forecasts}
\end{figure}

Realistic forecasts of day-ahead price and wind power production have been provided by Siemens Gamesa Renewable Energy for five consecutive months. Two years of synthetic forecasts are created for both price and wind production, by fitting the statistical characteristics of the forecast error of Siemens Gamesa  forecasts to a theoretical distribution, and then sampling from this distribution to generate new forecast errors with the same characteristics. Fig. \ref{fig:forecasts} provides a histogram of both original and generated forecast errors for both day-ahead market price and wind production forecasts.

Forecasts are produced by sampling forecast error values from the fitted distributions, and then adding these errors to the realized production data to create a forecast. The entire training and testing datasets thus consist of realized day-ahead prices, upward and downward balancing prices, and production for the wind farm in Roedsand, and forecasts for day-ahead prices and wind production, for $2019$ and $2020$.

\vspace{2mm}
\subsection{Models to be Compared}\label{sec:case_study-features}
\vspace{-1mm}
This numerical analysis compares several trained models with various combinations of (\textit{i}) linear policies architecture as introduced in Section \ref{Architecture} and (\textit{ii}) feature vectors. These trained models are compared to a deterministic (\textbf{Det.}) optimization approach as a base-case, and a benchmark with perfect information (\textbf{Hindsight}). In addition, we explore the impact of the length of the training period, ranging between $1$ and $12$ months, on the outcomes of different models.

We first compare models with \textbf{General Architecture (GA)} and \textbf{Hourly Architecture (HA)} for $\textbf{q}$-policies. Recall that $\textbf{q}$-policies take index $t$ in the \textbf{HA} model, which is not the case in the  \textbf{GA} model. Next, we compare the \textbf{GA} and \textbf{HA} in the cases where a single set of policies is defined over all values of prices, and a case where the policies are split into multiple \textbf{Price Domains (PD)}. When considering multiple price domains, we refer to models as \textbf{GA+PD} and \textbf{HA+PD}. Three price domains are chosen: one below the hydrogen price $\rho^{\rm H}\lambda^{\rm H}$, one above the $90$\% percentile  of realized day-ahead prices in the training data, and one in between. The hydrogen price is a natural threshold to separate the trading decisions because producing hydrogen is necessarily more profitable than selling power for day-ahead prices below the hydrogen price. The $90$\% percentile is chosen to reduce the influence of extreme outliers on the \textbf{q}-policies.

Additionally, we investigate three different types of feature vectors $\textbf{X}_t$. Firstly, the simplest \textbf{Reduced Feature (RF)} vector contains only the deterministic wind production forecast $\hat{P}^{\rm{W}}_t$, such that $\textbf{X}^{\rm RF}_t = \big[\hat{P}^{\rm{W}}_t, \lambda, 1 \big]$. Secondly, the \textbf{Augmented Feature (AF)} vector includes additional features, namely the aggregated onshore/offshore wind production forecasts in the two bidding zones DK1 and DK2, released by the Danish transmission system operator, Energinet, before the time of bidding, such that $\textbf{X}^{\rm AF}_t = \big[ \tilde{\textbf{X}}_t^{\rm AF}, \lambda, 1 \big]$ with $\tilde{\textbf{X}}_t^{\rm AF} = \big[
\hat{P}^{\rm W}_t, \hat{P}^{\rm on, DK1}_t, \hat{P}^{\rm on, DK2}_t, \hat{P}^{\rm off, DK1}_t,  \hat{P}^{\rm off, DK2}_t \big]$. This augmented feature vector may capture additional information related to the uncertainty sources at the cost of training additional $\textbf{q}$-policies  and requiring a larger amount of training data. Increasing the length of the historical dataset raises issues of stationarity of the environment, in particular of electricity prices. Therefore, a so-called \textbf{Forecast Model (FM)} feature vector is introduced, that accounts for the same additional features $\tilde{\textbf{X}}_t^{\rm AF}$, while minimizing the added complexity of the model. This is achieved by first training a separate feature-driven forecast model that provides a single improved wind power generation forecast feature $\tilde{X}^{\rm FM}_t =  \tilde{\textbf{q}}^{\rm FM} \big[ \tilde{\textbf{X}}_t^{\rm AF}, 1 \big]^\top \in \mathbb{R}$, where $\tilde{\textbf{q}}^{\rm FM}$ represents a vector of policies that is trained using a historical dataset. As the environment related to wind power generation is expected to be stationary, to a large extent, this feature-driven forecast model can be trained on a longer historical dataset. Then, the \textbf{FM} feature vector is defined as $\textbf{X}^{\rm FM}_t = \big[ \tilde{X}^{\rm FM}_t, \lambda, 1 \big]$. 

\vspace{2mm}
\subsection{Results}
\vspace{-1mm}
\begin{figure*}[t!]
    \centering
\subfloat[\label{fig:05_res_learned}]{\includegraphics[width=.3\linewidth]{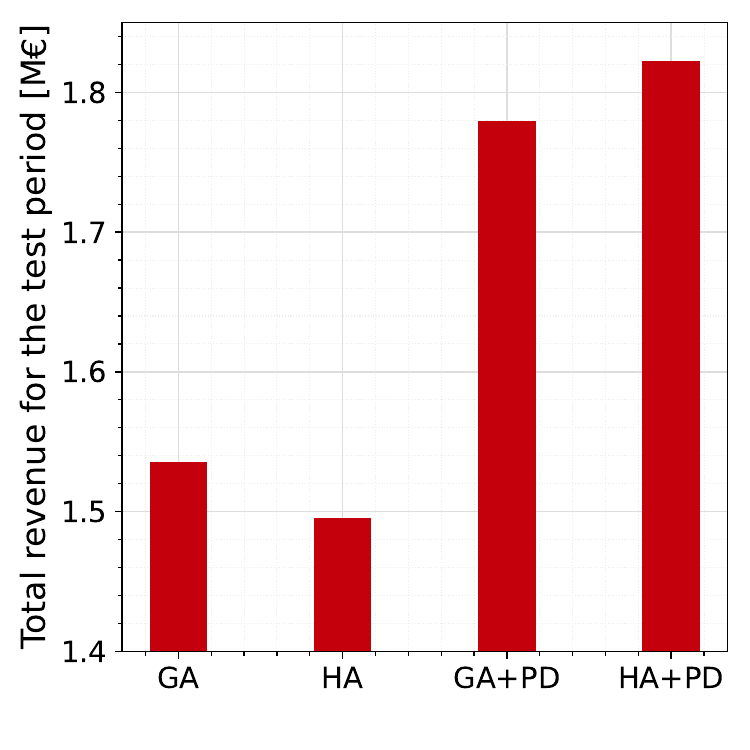}}
\subfloat[\label{fig:05_res_hapd}]{\includegraphics[width=.3\linewidth]{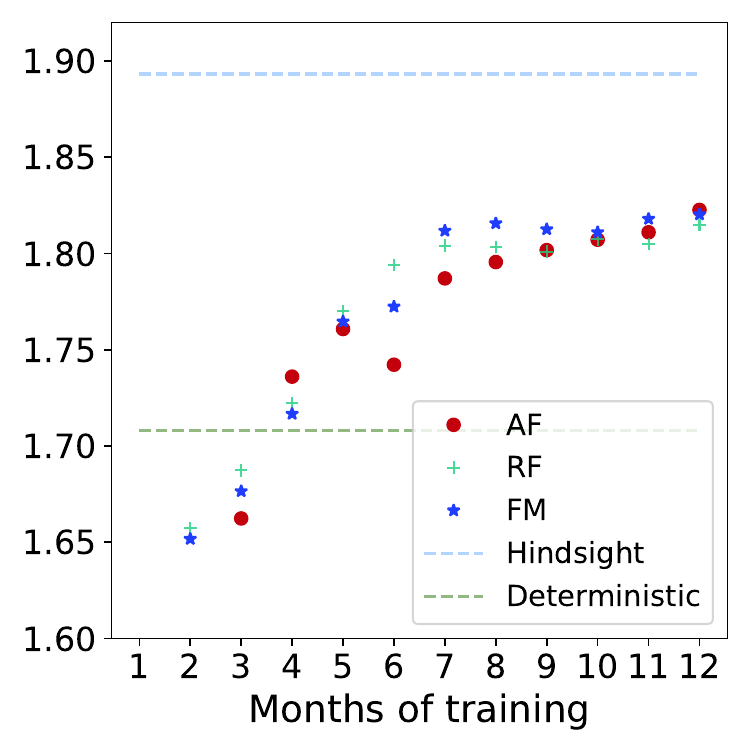}}
\subfloat[\label{fig:05_res_adj}]{\includegraphics[width=.3\linewidth]{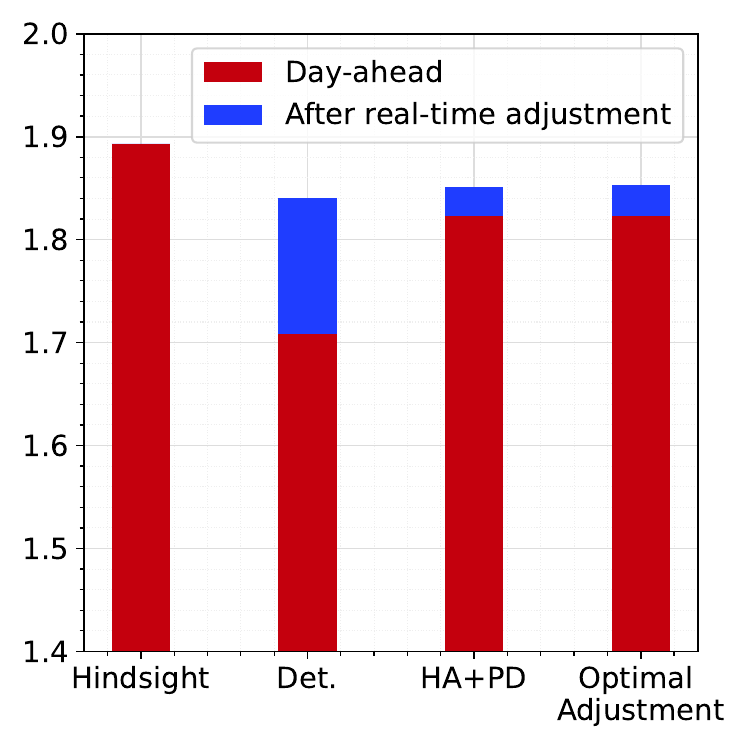}}
\caption{Ex-post profit of the hybrid power plant with (a) four policy architectures (\textbf{GA}, \textbf{HA}, \textbf{GA+PD}, \textbf{HA+PD}); (b) the \textbf{HA+PD} model for various training dataset lengths and available feature vectors (\textbf{AF}, \textbf{RF}, \textbf{FM}), compared to the \textbf{Deterministic} and \textbf{Hindsight} models; and (c) the \textbf{HA+PD} model before and after the real-time adjustment, compared to the deterministic (\textbf{Det.}) and  \textbf{Hindsight} model, and the \textbf{Optimal Adjustment} strategy.
}
\end{figure*}

The results presented are the profits generated by each model from electricity trades in the day-ahead market, hydrogen production, and settlements in real time, evaluated ex-post for the entire year $2020$.

Fig. \ref{fig:05_res_learned} illustrates the comparison between the different policy architectures, showing the best performing feature vector and training period for each architecture. This plot shows that introducing multiple price domains increases the performance radically for both types of architectures, resulting in the \textbf{HA+PD} model outperforming the \textbf{GA+PD} model with price domains included. Additionally, we note that the training period yielding the best results for the \textbf{GA} model is $5$ months, and the model performed within $99$\% of this performance after only $3$ months. With a training period of $3$ months, the \textbf{GA} model generates around $5$\% higher revenues than the \textbf{HA} model. This indicates that for time periods with a more non-stationary environment than the one used in the study, or with limited available historical data, the \textbf{GA} model might be a more appealing choice. Since Fig. \ref{fig:05_res_learned} shows that \textbf{HA+PD} is the model that outperforms others, we further analyze this model in the rest of this section.

Fig. \ref{fig:05_res_hapd} depicts the performances of \textbf{HA+PD} for different feature vectors and varying lengths of the training dataset, from one to $12$ months. First, we observe that the improvement of the model by having access to a training dataset longer than $7$ months is limited, irrespective of the feature vectors chosen. For example, the profit when models trained on a $12$-month dataset is only $1$-$2$\% higher than when trained on a $7$-month dataset. The second observation is that the model with the \textbf{FM} feature vector setting usually exhibits a better performance than the ones with the \textbf{AF} and \textbf{RF} feature vector settings, especially when the length of the training dataset is lower than $12$ months. However, these three models show a similar performance when the plant uses a dataset of $12$ months. The third observation is the performance of our models in comparison to the deterministic model, where the hybrid power plant makes day-ahead bidding decisions by using deterministic forecasts only. We notice that, by using a training dataset of at least $4$ months, our model outperforms the deterministic model. For example, when using the $12$-month dataset, the plant makes around $5.9\%$ more profit by using the \textbf{HA+PD} model with either of three feature vector settings. The final observation corresponds to the comparison of the \textbf{HA+PD} model with respect to the hindsight model, in which the plant has perfect information on the prices and wind generation. We observe that the profit earned by our proposed \textbf{HA+PD} model is only $3.7\%$ lower than the profit in hindsight, whereas it is $9.8\%$ for the deterministic model. For the rest of this section, we further analyze the best-performing model, i.e., the \textbf{HA+PD} model with a $12$-month training dataset and (\textbf{AF}) feature vector setting.




Fig. \ref{fig:05_res_adj} depicts the profit across various models before and after real-time adjustments. Obviously, the hindsight model does not need any real-time adjustment. The \textit{final} profit, i.e., after in the day-ahead stage plus real time, earned by the deterministic (\textbf{Det.}) and the proposed \textbf{HA+PD} model is computed using the rule-based adjustment strategy proposed in Section \ref{adjus}. This profit is compared to the final profit obtained by the HA+PD model with an \textit{optimal adjustment}, implemented as an optimization problem over $24$ hours with perfect foresight, which provides an upper bound on the benefits of any adjustment policy.
%
We observe that the the proposed real-time adjustment strategy achieves a higher increase in the profit of the deterministic model compared to that of the \textbf{HA+PD} model. Indeed, due to a less efficient day-ahead scheduling, the deterministic model has more potential for improvement in the real-time stage than the \textbf{HA+PD} model. However, the final profit of the deterministic model is still slightly lower than that of the proposed \textbf{HA+PD} model. Yet, it is important to note that its result relies on having access to accurate upward and downward balancing price forecasts for the the real-time adjustment strategy, which is not a straightforward task. As the final profits of the deterministic model are highly dependent on the performance of the real-time adjustment policy, in the absence of accurate price forecasts, the deterministic model risks achieving significantly lower final profits than the \textbf{HA+PD} model. Additionally, we observe that the proposed rule-base adjustment strategy achieves almost as much profit as the optimal adjustment one, under the assumption of accurate price forecasts. This shows that, owed to the proposed real-time adjustment strategy, the final profit of the \textbf{HA+PD} model gets even closer to the profit in hindsight, making it an even more attractive model.

\begin{figure}[t]
    \centering
\includegraphics[width=\linewidth]{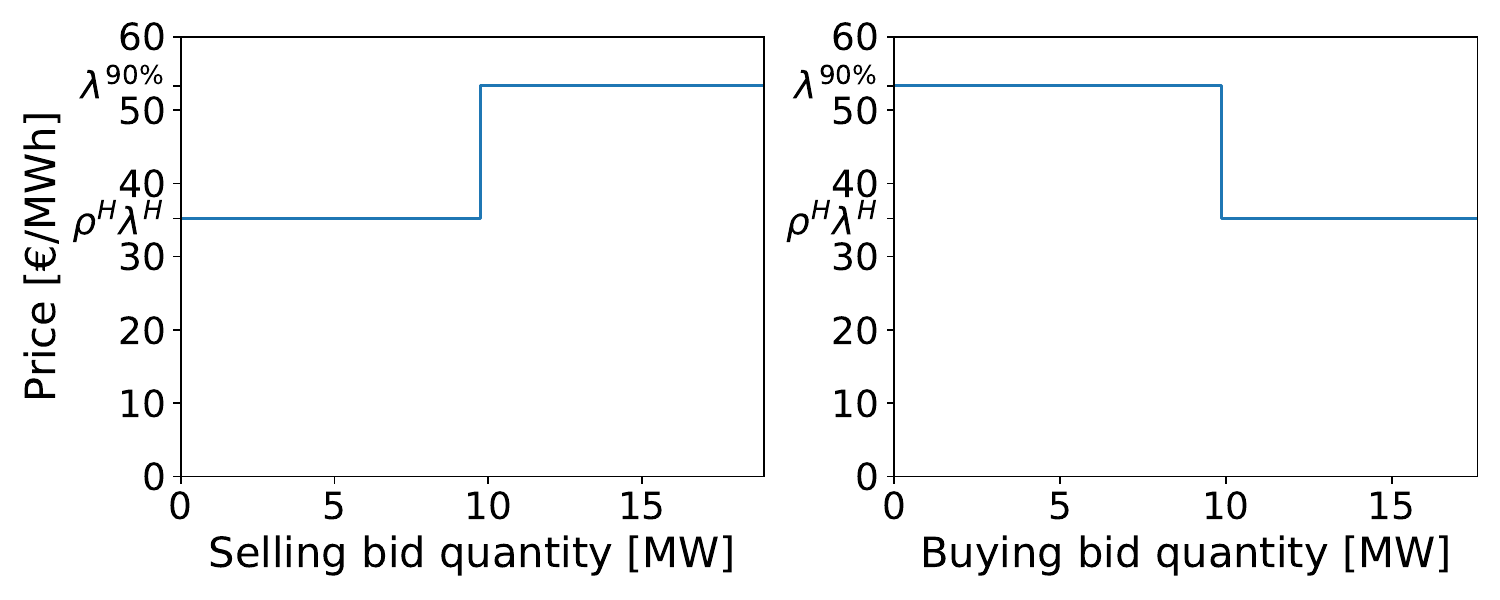}
\caption{Example of resulting price-quantity bidding curves in the day-ahead market, when the hybrid power plant buys \textbf{(right)} and sells \textbf{(left)} electricity. Recall $\rho^{\rm H}\lambda^{\rm H}$ is the hydrogen price, and $\lambda^{\rm 90\%}$ is the $90$\% percentile of realized day-ahead prices in the training data.}
\label{fig:example}
\vspace{-2mm}
\end{figure}

Finally, Fig. \eqref{fig:example} provides two examples of feature-driven bidding strategies in the day-ahead market, obtained by the \textbf{HA+PD} model. The left plot shows the bidding strategy in a representative hour during which the hybrid power plant sells power to the grid. In this hour, the plant sells the first $10$ MW at price $\rho^{\rm H}\lambda^{\rm H}$ and the rest at the $90$\% percentile of the realized day-ahead prices in the training data ($\lambda^{\rm 90\%}$). The right plot shows a representative hour during which the hybrid power plant buys power from the grid. The first $10$ MW is bought at price $\lambda^{\rm 90\%}$ and the rest at price $\rho^{\rm H}\lambda^{\rm H}$.

We applied a retraining procedure to all models with a monthly retraining schedule. This retraining procedure is expected to improve the performance of the algorithm over time, for instance, adapting to seasonal wind patterns and changes in electricity prices distribution. However, over the testing period considered, retraining provided minimal improvements, and is therefore not considered adequately to draw any conclusions about its impact.

\vspace{5mm}
\section{Conclusion}\label{sec:conclusion}
This paper explains how hybrid power plants with co-located wind turbines and an electrolyzer can implement efficient feature-driven models to learn from historical data and make informed day-ahead trading decisions. The proposed feature-driven model derive trading (selling or buying) decisions in the day-ahead electricity market, as well as a hydrogen production schedule for the next day, fulfilling the minimum daily hydrogen quota. This is a pragmatic solution, which properly accounts for wind power and price uncertainty without the need to generate probabilistic forecasts or solve complex stochastic optimization problems. Our numerical analysis shows that the proposed feature-driven models outperform the deterministic model, and require less adjustment in  real time. In addition, they result in a final profit which is close to that in hindsight. 

For future work, it is of interest to explore the performance of the model in a non-stationary environment. This could be tackled by developing online decision-making methods, similar to the one proposed in \cite{miguel} for wind power trading only. Another valuable potential of implementing an online decision-making method is an expansion of the adjustment model to account for the uncertainty in the prediction of the balancing price and thereby improve the value of real-time adjustment. It is also interesting to include more complex physical characteristics of the electrolyzer, such as non-linear efficiency and degradation curves, and add more relevant assets to the hybrid power plant, such as battery and hydrogen storage. 

\vspace{5mm}
\section*{Acknowledgement}
We would like to thank Siemens Gamesa Renewable Energy for providing us with historical forecast data. We also thank the Danish Energy Technology Development and Demonstration Programme (EUDP) for supporting this research through HOMEY (Grant number: 64021-7010) and ViPES2X (Grant number: 640222-496237) projects. 
\vspace{5mm}







\bibliographystyle{IEEEtran}
\bibliography{reference}

\vfill

\end{document}